\documentclass[a4paper]{aa}
\usepackage{graphics,txfonts,amsfonts}

\usepackage{natbib}
\bibpunct[, ]{(}{)}{;}{a}{}{,}

\newcommand{\gbar}{\overline{g}}
\newcommand{\Gbar}{\overline{G}}
\newcommand{\syn}{{\sc synmast}}
\newcommand{\teff}{$T_{\rm eff}$}
\newcommand{\logg}{$\log g$}
\newcommand{\kms}{km\,s$^{-1}$}

\newcommand{\figps}[1]{\resizebox{\hsize}{!}{\rotatebox{0}{\includegraphics{#1}}}}

\newcommand{\firps}[1]{\resizebox{\hsize}{!}{\rotatebox{90}{\includegraphics{#1}}}}
\newcommand{\fifps}[2]{\centering\resizebox{#1}{!}{\includegraphics{#2}}}

\newcommand{\beq}{\begin{equation}}
\newcommand{\eeq}{\end{equation}}

\begin{document}

\title{Least-squares deconvolution of the stellar intensity \\ and polarization spectra}

\author{O. Kochukhov
   \and V. Makaganiuk
   \and N. Piskunov}

\offprints{O. Kochukhov, \\ \email{oleg.kochukhov@fysast.uu.se}}

\institute{Department of Physics and Astronomy, Uppsala University, Box 516, Uppsala SE-751 20, Sweden}

\date{Received 19 July 2010 / Accepted 31 August 2010}

\abstract%
{
Least-squares deconvolution (LSD) is a powerful method of extracting high-precision average line profiles from the stellar intensity and polarization spectra. This technique is widely used for detection, characterization, and detailed mapping of the temperature, magnetic, and chemical abundance structures on the surfaces of stars.
}
{
Despite its common usage, the LSD method is poorly documented and has never been tested with realistic synthetic spectra. In this study we revisit the key assumptions of the LSD technique, clarify its numerical implementation, discuss possible improvements and give recommendations of how to make LSD results understandable and reproducible. We also address the problem of interpretation of the moments and shapes of the LSD profiles in terms of physical parameters.
}
{
We have developed an improved, multiprofile version of LSD (iLSD) and have extended the deconvolution procedure to linear polarization analysis taking into account anomalous Zeeman splitting of spectral lines. The iLSD method is applied to the theoretical Stokes parameter spectra computed for a wide wavelength interval containing all relevant spectral lines. We test various methods of interpreting the mean profiles, investigating how coarse approximations of the multiline technique translate into errors of the derived parameters.
}
{
We find that, generally, the Stokes parameter LSD profiles do not behave as a real spectral line with respect to the variation of magnetic field and elemental abundance. This problem is especially prominent for the Stokes $I$ (intensity) variation with abundance and Stokes $Q$ (linear polarization) variation with magnetic field. At the same time, the Stokes $V$ (circular polarization) LSD spectra closely resemble the profile of a properly chosen synthetic line for the magnetic field strength up to 1~kG. The longitudinal field estimated from the Stokes $V$ LSD profile is accurate to within 10\% for the field strength below 5~kG and to within a few percent for the fields weaker than 1 kG. Our iLSD technique offers clear advantages over the standard LSD method in the individual analysis of different chemical elements.
}
{
We conclude that the usual method of interpreting the LSD profiles by assuming that they are equivalent to a real spectral line gives satisfactory results only in a limited parameter range and thus should be applied with caution. A more trustworthy approach is to abandon the single-line approximation of the average profiles and apply LSD consistently to observations and synthetic spectra.
}

\maketitle

\section{Introduction}
\label{intro}

Many modern echelle spectrographs allow one to record high-resolution stellar spectra for the entire wavelength range spanning from the near-UV to near-IR (3500--10000~\AA). These data contain an enormous amount of information about the physical conditions, dynamics, and chemistry of the stellar surface layers. However, often only a small number of spectral features is used in astrophysical analyses that are based on these observations. This is justified when one is interested in specific diagnostic lines or in chemical elements with sparse spectra. On the other hand, there are many processes that produce a very similar imprint on many spectral lines. For instance, the radial velocity shifts owing to the orbital motion in a binary system, non-radial stellar pulsations, and cool spots on the surfaces of late-type active stars affect all spectral lines in a similar way. Thus, one can employ a \textit{multiline technique} to extract common physical information from many spectral lines and determine an average profile with a high signal-to-noise ($S/N$) ratio.

Early applications of the multiline spectrum analysis focused on the detection of the weak magnetic field in cool stars using circular polarization observations. Direct detection of complex and generally weak polarization signatures of the dynamo-generated magnetic fields in individual spectral lines requires a $S/N\approx1000$ \citep{donati:1992}, which is impossible to reach for all but the brightest objects. \citet{semel:1989} and \citet{semel:1996} suggested to average polarization measurements of many spectral lines to detect and analyze magnetic fields with polarization observations of a modest quality. \citet{donati:1997} refined this simple \textit{line addition} procedure into the \textit{least-squares deconvolution} (LSD) method. This technique assumes that all spectral lines have the same profile, scaled by a certain factor, and that overlapping lines add up linearly. These assumptions lead to a highly simplified description of the intensity and circular polarization observations in terms of the convolution of a known line mask with an unknown profile. With this approximation one can reconstruct an average line shape, the LSD profile, which is formally characterized by an extremely high $S/N$ ratio.

The LSD analysis of the circular polarization observations has been very successful in detecting weak magnetic fields in stars over the entire Hertzsprung-Russell diagram \citep[e.g.,][]{petit:2008b,lignieres:2009,auriere:2009,alecian:2008,donati:2008b,morin:2008}. The LSD was also extended to the Stokes $Q$ and $U$ spectra by \citet{wade:2000b} with the aim to characterize weak linear polarization signatures in magnetic Ap stars. Outside the research area of stellar activity and magnetism LSD was applied to disentangle composite spectra of multiple stars and measure their orbital radial velocity variation \citep{hareter:2008}. A modification of the LSD method was also employed in the search for stellar differential rotation through the line profile analysis \citep{reiners:2003}.

Many studies went further than using LSD as a simple magnetic field and starspot detection tool, attempting to interpret the LSD profiles and extract various physical information. These applications range from a relatively simple determination of the longitudinal magnetic field from the LSD Stokes $V$ \citep{auriere:2007} and of the net linear polarization from the LSD Stokes $QU$ data \citep{wade:2000} to numerous sophisticated Doppler and Zeeman-Doppler imaging studies. In the latter investigations the time series of the LSD profiles were interpreted under different physical assumptions, with regularized inversion techniques to produce magnetic \citep[e.g.,][]{donati:2003}, chemical \citep{folsom:2008}, and brightness \citep[e.g.,][]{jeffers:2007} maps of stellar surface. All these studies implicitly assumed that the LSD profile behaves similar to a real, isolated spectral line with average parameters.

The proliferation of the studies that use LSD outside its original validity range and assign certain meaning to the LSD profiles without theoretical justification or numerical tests casts doubt on some of the results obtained with this method. Few papers attempted to clarify or improve the multiline technique. \citet{barnes:2004} studied mathematical aspects of the spectral deconvolution procedure. \citet{sennhauser:2009a} attempted to improve the LSD description of the intensity spectrum by accounting for nonlinearity in blended profiles. \citet{semel:2009} and \citet{ramirez-velez:2010} tested alternative multiline Zeeman signature methods (similar but not identical to LSD). 

None of the previous studies came close to characterizing the LSD technique and testing the reliability of the LSD profile interpretation for a realistic situation of blended intensity and polarization profiles of thousands of spectral lines. This task was beyond the capabilities of a physically complete spectrum synthesis for spotted magnetic stars, which was typically limited to a few dozen spectral lines in several short wavelength intervals \citep[e.g.,][]{kochukhov:2004d}. Recent progress in numerical techniques of the polarized spectrum synthesis and continuous expansion of computer resources has allowed us to overcome these limitations. In this paper we present the first comprehensive, direct numerical tests of the LSD technique, complemented with an improvement and clarification of the LSD method itself. We compute four Stokes parameter stellar spectra in a wide wavelength region, including all relevant metal absorption lines. We then apply LSD to these synthetic spectra and compare quantities inferred from the LSD profiles with the input parameters adopted for polarized spectrum synthesis. This work allows us to reach general conclusions about the validity of different applications of LSD in stellar physics.

Our paper is structured as follows. In Sect.~\ref{lsd} we revisit the general principle and basic assumptions of the LSD technique. Then we describe numerical implementation of the deconvolution procedure in our LSD code and consider possible improvements of the least-squares deconvolution technique. We also address the question of normalization of the LSD profiles and outline several methods of their interpretation. Section~\ref{tests} is devoted to numerical tests of the LSD technique with the help of theoretical polarized spectrum synthesis. We start with a brief description of the magnetic spectrum synthesis code that we use to simulate realistic spectra in four Stokes parameters. We examine the behavior of the LSD profile shapes and of the integral quantities that can be inferred from the average profiles in different Stokes parameters (longitudinal magnetic field, net linear polarization, equivalent width). The paper concludes with the summary of the main results of our investigation in Sect.~\ref{conc}.

\section{Least-squares deconvolution}
\label{lsd}

\subsection{General principle}

We consider a situation where each line in the intensity or polarization spectrum of a star can be approximately described by the same line shape, scaled by a certain factor. If we further assume that overlapping line profiles add up linearly, the entire spectrum $Y(v)$ can be described as a sum of scaled and shifted identical profiles $Z(v)$
\beq
Y(v) = \sum_i w_i \delta(v-v_i) Z(v_i),
\label{eq:1}
\eeq
where $Y$ is a model residual intensity, $1-I/I_{c}$, or normalized polarization spectrum, e.g. $V/I_{c}$. The position in the velocity space $v_i=c\Delta\lambda_i/\lambda_i$ is associated with a wavelength shift $\Delta\lambda_i$ from the central wavelength of the $i$-th line. The relative contribution of each spectral line is given by the weight $w_i$.

Following \citet{donati:1997}, Eq.~(\ref{eq:1}) can be expressed as a convolution of the line pattern function
\beq
M(v) = \sum_i w_i \delta(v-v_i)
\eeq
and the mean profile $Z(v)$
\beq
Y = M * Z,
\label{eq:3}
\eeq
or, equivalently, as a matrix multiplication
\beq
\mathbf{Y}=\mathbf{M} \cdot \mathbf{Z},
\label{eq:4}
\eeq
where $\mathbf{Y}$ is an $n$-element model spectrum vector, $\mathbf{Z}$ is an $m$-element common profile spanning a certain range in velocity, and $\mathbf{M}$ is a $n\times m$ line pattern matrix containing information on the line positions and relative strengths.
In typical applications $m$ might be on the order of $10^2$ or $10^3$, while $n$ might be on the order of $10^5$ or $10^6$. 

The scope of the LSD technique is to solve an inverse problem corresponding to Eq.~(\ref{eq:3}) or (\ref{eq:4}). Namely, to estimate the mean line profile $Z$ for a given line pattern information $M$, observed spectrum $Y^{\rm o}$ and associated statistical uncertainty $\sigma$. In general, this approach is justified as long as the average profile can be interpreted meaningfully in terms of useful physical quantities, and its uncertainty $\sigma_Z$ is much smaller than the error of the input spectrum, $\sigma_Z\ll\sigma$.

An average line profile can be also estimated by the direct superposing, weighting, and averaging a large number of short pieces of spectrum. This approach would yield results identical to LSD if all lines were free of blends. In reality, absorption lines in stellar spectra are rarely isolated. For this reason direct averaging recovers a profile that is systematically distorted by overlapping lines, whereas LSD is able to provide a clean mean profile.

Note that in its application to the Stokes $I$ spectrum the LSD method is similar to the broadening function (BF) analysis technique developed by \citet{rucinski:1992}. The essential difference is that the BF technique starts from an unbroadened observed or synthetic template spectrum rather than a superposition of $\delta$-functions employed by LSD. The requirement of using a realistic input template imposes certain limitations in practical applications of the BF method and does not allow a straightforward extension of this technique to the multiline polarization analysis. On the other hand, the BF method provides a more precise approximation of the intensity spectrum because it does not rely on the assumption of a linear addition of blended lines in the template spectrum.

\subsection{Physical assumptions}

The simplified description of the stellar spectrum adopted by LSD seems to be very coarse and is obviously invalid in some extreme cases. For instance, there is very little similarity of the spectral line profiles when the magnetic field strength is so large that individual Zeeman splitting patterns of spectral lines are resolved. But in some other cases, such as unsaturated lines in weak magnetic field, the LSD assumptions are not so bad. The goal of this section is to discuss which \textit{physical approximations} correspond to the mathematical expressions employed in the LSD method and to qualitatively estimate the validity range of these assumptions.

\subsubsection{Intensity}
\label{inten}

Let us start from the assumption of the line profile similarity in non-magnetic case. It is known that if the lines are unsaturated, the local residual intensity profile emerging from a stellar atmosphere is proportional to the line absorption coefficient $\kappa_\ell$ at a certain depth
\beq
1 - X^I_{\ell}(\lambda)/X^I_{c} \propto \kappa_\ell(\lambda)
\label{eq:5}
\eeq
and, therefore, can be represented as an average local profile $Z^I_{\rm loc}$ scaled in depth by the corresponding central line depression $d_{\rm loc}$ and in width by the wavelength $\lambda$
\beq
1- X^I_{\ell}(v)/X^I_{c}= d_{\rm loc} Z^I_{\rm loc} (v).
\eeq
Thus, for LSD to be applicable to the local Stokes $I$ spectrum $X^I$, spectral lines must all have a similar shape of the absorption coefficient $\kappa_\ell$ \textit{and} be relatively weak. The former condition suggests that the multiline technique should avoid spectral features with an unusual line shape, such as the \ion{H}{i} and \ion{He}{ii} lines in hot stars or metal lines exhibiting strong damping wings in cool stars. 

\begin{figure}[!t]
\fifps{8cm}{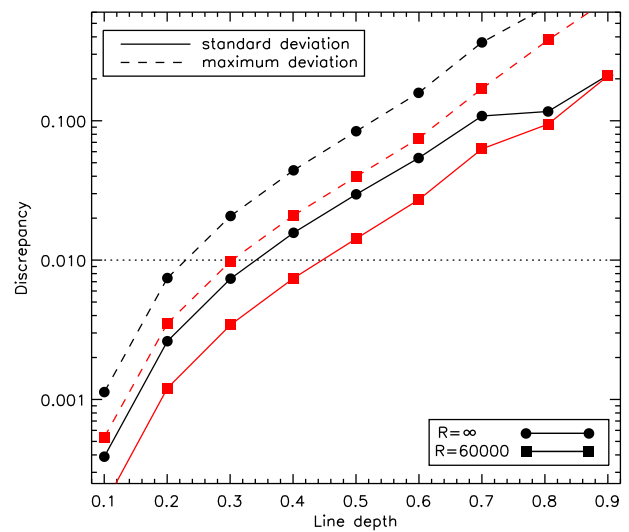}
\caption{Discrepancy between scaled profile of the weak \ion{Fe}{i} spectral line and detailed spectrum synthesis calculation for the line of given central depth. The discrepancy is characterized with the standard deviation (\textit{solid line}) and maximum difference (\textit{dashed line}). Results are shown for an unbroadened spectrum (\textit{circles}) and for the resolving power $R=60\,000$ (\textit{squares}).}
\label{fig:line_scale}
\end{figure}

\begin{figure}[!t]
\fifps{8cm}{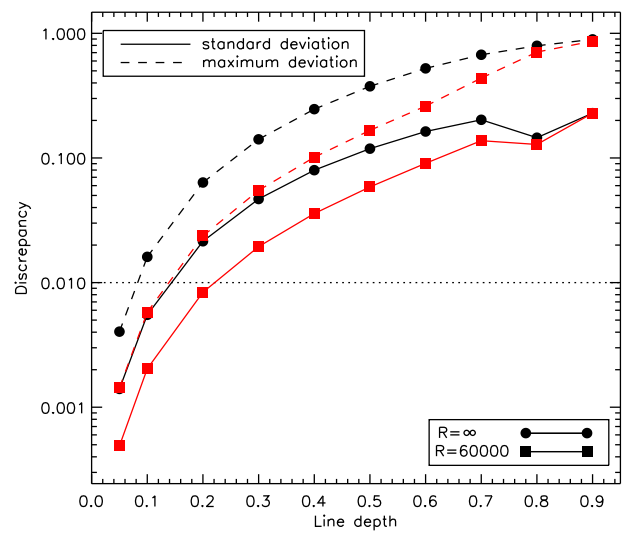}
\caption{Same as Fig.~\ref{fig:line_scale} for the discrepancy between linearly added profiles of two identical \ion{Fe}{i} lines of a given central depth and detailed spectrum synthesis calculation.}
\label{fig:line_blend}
\end{figure}

A simple assessment of the validity of the line shape self-similarity condition can be obtained by considering realistic theoretical profiles of typical spectral lines of a different strength. With the spectrum synthesis technique described in more detail below (Sect.~\ref{synt}), we calculated local profiles of fictitious \ion{Fe}{i} line with representative parameters and oscillator strength adjusted to yield central depths from 5\% to 90\% of the continuum intensity. The discrepancy between the real line profile and a scaled version of the weakest line is illustrated in Fig.~\ref{fig:line_scale} for an infinite resolving power and for $R=60\,000$. Keeping in mind the typical observational uncertainty of $\sim$\,$10^{-2}$ of the modern echelle spectra of bright stars, one can conclude that an identical line shape assumption of the LSD of Stokes $I$ is appropriate only for lines weaker than $\approx$\,40\%. These results were obtained with a model atmosphere typical of a mid-A star (\teff\,=\,9000~K, \logg\,=\,4.0), but a calculation with a solar model atmosphere yielded very similar results.

An analogous calculation can give us an idea about the accuracy of the linear addition of the blend components. This approximation is evidently valid for the spectral lines described by Eq.~(\ref{eq:5}), but it becomes increasingly unrealistic for stronger lines. Figure~\ref{fig:line_blend} illustrates the discrepancy between detailed spectrum synthesis of the blend consisting of two identical \ion{Fe}{i} lines with a given strength and simple arithmetic addition of the line depressions corresponding to the individual blend components. This figure shows that the accuracy of the linear line-adding remains at the level of $\le$\,10$^{-2}$ if the residual intensity of the blend components is $\la$\,20\%.

Until now we were examining the behavior of the \textit{local} intensity profiles which, of course, are not directly observable in real stars. In reality, the LSD method approximates the \textit{disk-integrated} stellar spectra. Integration over all points $M$ on the visible stellar disk takes into account the distribution of the continuum brightness $b_M$ and rotational Doppler shift $v_M$
\beq
1-I(v)/I_{c} = \int_S b_M d_{\rm loc} Z^I_{\rm loc} (M,v-v_M)\mathrm{d}S \approx d Z^I (v).
\eeq
Here $d$ is the line depth in the disk-integrated spectrum and the second equality is approximate because we have assumed that the center-to-limb variation of the normalized local line profiles of all lines can be described by the same linear law, independent of the wavelength and atomic line parameters. This assumption is not a particularly restrictive one for the majority of normal stars. However, one can find examples, such as the spectra of rapidly rotating early-type stars, where the line-to-line difference in the center-to-limb variation is emphasized by an inhomogeneous temperature distribution and can be readily detected in the stellar flux spectrum \citep{takeda:2008}.

\subsubsection{Polarization}
\label{polar}

The characteristic splitting of the Zeeman components of a spectral line is
\beq
\Delta\lambda_{\rm B} = \frac{\lambda_c^2 e_0 B}{4\pi m_{\rm e} c^2},
\eeq
where $\lambda_c$ is the wavelength of the unperturbed line, $B$ is the local field strength, and the other variables have their usual meaning. When the magnetic splitting is much smaller than the intrinsic width of the spectral lines determined by the thermal Doppler broadening $\Delta\lambda_{\rm D}$, one can derive approximate expressions relating intensity (Stokes $I$) with the Stokes parameters characterizing circular (Stokes $V$) and linear (Stokes $QU$) polarization \citep[e.g.,][]{polarization:2004}. 

In the \textit{weak field limit}, formally defined by the condition $\Delta\lambda_{\rm B} \ll \Delta\lambda_{\rm D}$, any spectral line obeys
\beq
V(\lambda) \propto \gbar \frac{\Delta\lambda_{\rm B}}{\Delta\lambda_{\rm D}} \frac{\partial I}{\partial\lambda}.
\eeq
The effective Land\'e factor $\gbar$ entering this expression is given by
\beq
\gbar = \frac{1}{2} (g_1 + g_2) + \frac{1}{4} (g_1 - g_2) r,
\eeq
where
\beq
r = J_1 (J_1 + 1) - J_2 (J_2 + 1).
\eeq
This relation between Stokes $V$ and the first derivative of Stokes $I$ is valid for lines of any strength.
The corresponding weak-field relation for the linear polarization,
\beq
Q, U \propto \Gbar \left(\frac{\Delta\lambda_{\rm B}}{\Delta\lambda_{\rm D}} \right)^2 \frac{\partial^2 I}{\partial\lambda^2}
\eeq
with
\beq
\Gbar = \gbar^2 - \frac{1}{80} (g_1 - g_2)^2 (16 s - 7 r^2 - 4)
\eeq
and
\beq
s = J_1 (J_1 + 1) + J_2 (J_2 + 1),
\eeq
is derived using a set of more restrictive assumptions \citep[see][]{polarization:2004}, essentially equivalent to the weak line approximation of Sect.~\ref{inten}.

Following the same arguments as given above for the transformation from the local to disk-integrated intensity spectra and taking into account that $\Delta\lambda_{\rm B}/\Delta\lambda_{\rm D} \sim \lambda_{\rm c}$, we conclude that the disk-integrated Stokes $V$ profiles can be represented by a common profile scaled by a factor $w_V = d \lambda_{\rm c} \gbar$:
\beq
V(v)/I_{\rm c} \approx w_V Z^V(v).
\label{vsim}
\eeq
The linear polarization line shapes are approximated with
\beq
Q(v)/I_{\rm c} \approx w_Q Z^Q(v),
\label{qsim}
\eeq
where the linear polarization line weight $w_Q=d \lambda^2_{\rm c} \Gbar$.

\citet{wade:2000b} established a similar relation for the scaling of the linear polarization profiles, but they used $\gbar^2$ instead of $\Gbar$ in the line weight $w_Q$. This is equivalent to assuming that all spectral lines have a simple triplet splitting pattern. In general, $\Gbar < \gbar^2$ for the anomalous Zeeman splitting, and sometimes $\Gbar < 0$. In Sect.~\ref{tests} we demonstrate that our generalization of the LSD linear polarization profile scaling factor yields a better approximation of the Stokes parameter spectra obtained with detailed spectrum synthesis calculations.

We emphasize that the self-similarity of the polarization profiles asserted by Eqs.~(\ref{vsim}) and (\ref{qsim}) is derived assuming both the weak field regime for the magnetic splitting and the weak line approximation for the line formation (implicitly for Stokes $V$ and explicitly for Stokes $QU$). The validity of the second assumption was assessed in Sect.~\ref{inten}. On the other hand, the weak field approximation does not have a well-defined validity range. Most authors agree that it is fairly accurate only for the magnetic field strengths below $\sim$\,1~kG. However, this approximation is also widely used for the LSD analysis of multi-kG fields of Ap stars \citep{wade:2000}. 

We suggest that it makes little practical sense to assign a general ``validity range'' to the LSD polarization profiles and quantities derived from them. Instead we aim to quantify the LSD profile behavior numerically and assess the robustness of their interpretation in the context of specific applications, taking into consideration the quality of observational data.

\subsection{Numerical implementation}
\label{num}

We follow \citet{donati:1997} in representing the LSD model spectrum as a matrix multiplication described by Eq.~(\ref{eq:4}). We use linear interpolation to project the mean profile $Z(v)$ onto an arbitrary wavelength grid of observations. Then a spectral line with a weight $w_l$ and central wavelength $\lambda_l$ contributes along a bidiagonal of the line mask matrix $\mathbf{M}$
\beq
\begin{array}{lcl}
M_{i,j}  & = & w_l (v_{j+1} - v_i) / (v_{j+1} - v_j) \\
M_{i,j+1} & = & w_l (v_{i} - v_j) / (v_{j+1} - v_j)
\end{array}
\eeq
with
\beq
v_i = c (\lambda_i-\lambda_l)/\lambda_l \mathrm{~and~} v_j \le v_i \le v_{j+1}.
\eeq
Here index $i$ corresponds to the wavelength space of observations, whereas index $j$ runs over all velocity points of the LSD profile. 

The problem of finding the best-fitting mean profile $Z(v)$ for a given line mask $\mathbf{M}$ and observations $\mathbf{Y}^{\rm o}$ is equivalent to minimizing the $\chi^2$ function
\beq
\chi^2 = \left(\mathbf{Y}^{\rm o} - \mathbf{M} \cdot \mathbf{Z}\right)^T \cdot \mathbf{S}^2 \cdot
\left(\mathbf{Y}^{\rm o} - \mathbf{M} \cdot \mathbf{Z}\right) \to \mathrm{min},
\label{eq:lsq}
\eeq
where $\mathbf{S}$ is the square diagonal matrix containing inverse variance, $S_{ii}=1/\sigma_i$, for individual pixels in the observed spectrum.

Since the model describing observations is linear, the least-squares solution of the $\chi^2$ problem formulated in Eq.~(\ref{eq:lsq}) is given by 
\beq
\mathbf{Z} = (\mathbf{M}^T \cdot \mathbf{S}^2 \cdot \mathbf{M})^{-1} \cdot \mathbf{M}^T \cdot \mathbf{S}^2 \cdot \mathbf{Y}^{\rm o}.
\label{eq:lsd}
\eeq
The $\mathbf{M}^T \cdot \mathbf{S}^2 \cdot \mathbf{Y}^{\rm o}$ term in Eq.~(\ref{eq:lsd}) represents a weighted cross-correlation between the line mask and the observed spectrum. The left-hand part is the inverse of autocorrelation matrix $\mathbf{M}^T \cdot \mathbf{S}^2 \cdot \mathbf{M}$, which effectively deconvolves the cross-correlation vector and provides formal uncertainty estimates of the LSD profile through the diagonal elements of $(\mathbf{M}^T \cdot \mathbf{S}^2 \cdot \mathbf{M})^{-1}$.

We have implemented the numerical technique described above, along with the improvements related to the linear polarization treatment (Sect.~\ref{polar}), multiple mean profile analysis (Sect.~\ref{multi}), and regularized deconvolution (Sect.~\ref{reg}), in an \textit{improved least-squares deconvolution} (iLSD) Fortran code. On input it reads intensity or polarization observations and the corresponding error bars. Several sets of line weights are defined explicitly in another input file. Then, the code fills the line pattern matrix $\mathbf{M}$ for a prescribed mean profile velocity grid and evaluates $Z(v)$ using Eq.~(\ref{eq:lsd}). The inverse of the autocorrelation matrix is calculated with the help of LU-decomposition \citep{press:1992}. The error bars of the resulting LSD profile are scaled to ensure that the reduced chi-square is equal to 1. In our implementation of the LSD technique we consider only those pixels of the observed spectrum that contain the contribution of at least one spectral line.

As discussed in Sect.~\ref{inten}, the coarse LSD model spectrum calculated with Eq.~(\ref{eq:4}) significantly overestimates the intensity of strong blends, sometimes leading to unphysical results. \citet{donati:1997} mention that they ensure that ``the sum of normalized central depths over nearby lines never exceeds 1''. However, it is not clear how this constraint can be implemented within the linear LSD approach. G.~Wade (private communication) informs us that this modification is not used in the practical applications of Donati's LSD code. Thus, we also solve the least-squares problem defined by Eq.~(\ref{eq:lsq}) without imposing any additional constraints.

\subsection{Improvements of deconvolution technique}
\label{improve}

\subsubsection{Multiprofile LSD}
\label{multi}

A straightforward generalization of the single-profile LSD technique can be obtained by postulating that each line in the stellar spectrum is represented by a superposition of $N$ different scaled mean profiles:
\beq
\mathbf{Y}=\sum_{k=1}^N \mathbf{M}_k \cdot \mathbf{Z}_k = \mathbf{M}^\prime \cdot \mathbf{Z}^\prime.
\eeq
In that case one can still use the matrix formulation of the LSD technique, but with a composite line pattern matrix $\mathbf{M}^\prime$ with the dimensions $n \times (m\cdot N)$. The matrix $\mathbf{M}^\prime$ is formed by concatenating matrices $\mathbf{M}_k$ that have the same structure, but a different magnitude of the non-zero elements corresponding to $N$ different sets of line weights. All $N$ average profiles, concatenated in the composite profile $\mathbf{Z}^\prime$, are reconstructed simultaneously with the matrix inversion technique described in Sect.~\ref{num}.

There are several advantages of using multiprofile LSD. One can, to some extent, account for the line shape difference between weak and strong lines or recover several mean profiles if the stellar spectrum contains two or more systems of lines formed under sufficiently different physical conditions (e.g., a composite spectrum of a binary star).

In the studies of magnetic chemically peculiar stars, which often exhibit diverse distributions of chemical elements, one has to consider LSD profiles of individual elements \citep{wade:2000b,folsom:2008}. Using conventional single-profile LSD for a given element often introduces significant distortions in the wings of the corresponding mean profile owing to neglected blending by lines of other elements. In contrast, with the multiprofile least-squares deconvolution we are able to account for the background lines blending the element of interest. We have implemented multiprofile LSD in our iLSD code and apply it extensively in the numerical tests presented in Sect.~\ref{tests}.

\subsubsection{Regularized deconvolution}
\label{reg}

\begin{figure}[!t]
\figps{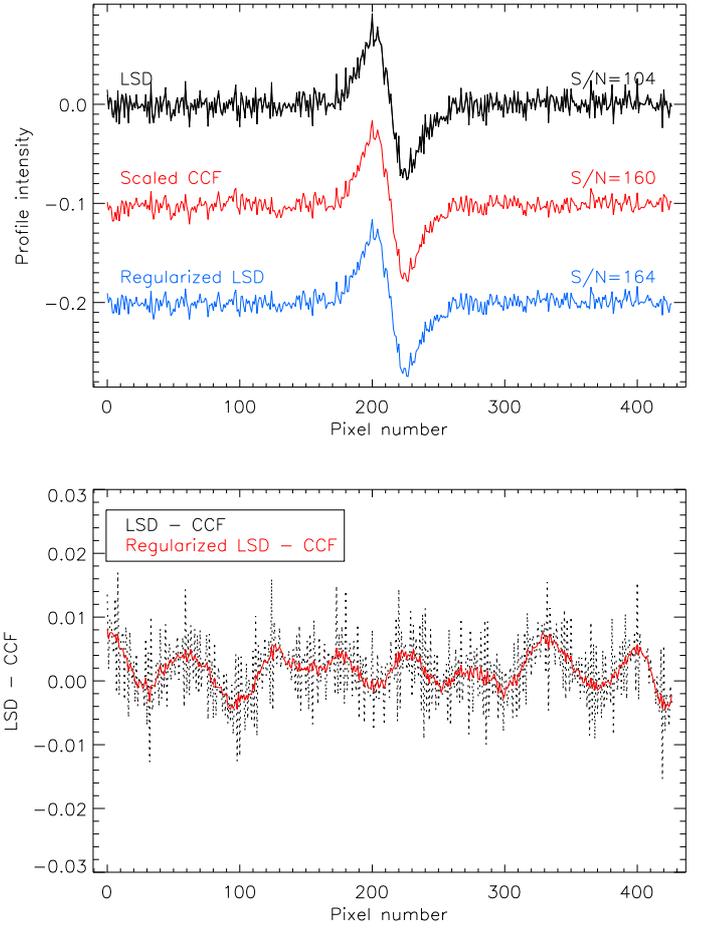}
\caption{{\it Top panel:} Comparison between the usual LSD profile, the scaled cross-correlation function and the LSD profile reconstructed using regularization. These profiles are obtained from a simulated S/N=15 spectrum, containing 500 randomly distributed scaled lines with a characteristic Stokes $V$ S-shaped signature. Profiles are shifted vertically for display purpose. The indicated signal-to-noise ratio is inferred from the pixel scatter in the regions on both sides of the profile. {\it Bottom panel:} The difference between LSD profiles and the corresponding CCF.}
\label{fig:reg}
\end{figure}

With the multiline technique described by Eq.~(\ref{eq:lsd}) one derives the mean profile in two steps. First, a cross-correlation function (CCF) is obtained and, second, it is multiplied by an inverse of the autocorrelation matrix to remove the spurious pattern arising from possible regularities in the line distribution. However, the latter operation suffers from a well-known problem of noise amplification, which is common to all applications of direct deconvolution. In the present case the second step systematically increases the noise in the mean profile compared to CCF independently of the numerical technique used to invert the autocorrelation matrix. This is illustrated in Fig.~\ref{fig:reg}, which shows the reconstruction of the mean profile with a typical Stokes $V$ signature from 500 superimposed, scaled copies of this profile that are randomly distributed in wavelength. Random noise was added to simulate observational data with a signal-to-noise ratio of 15.

\begin{figure*}[!t]
\centering
{\resizebox{16cm}{!}{\rotatebox{90}{\includegraphics{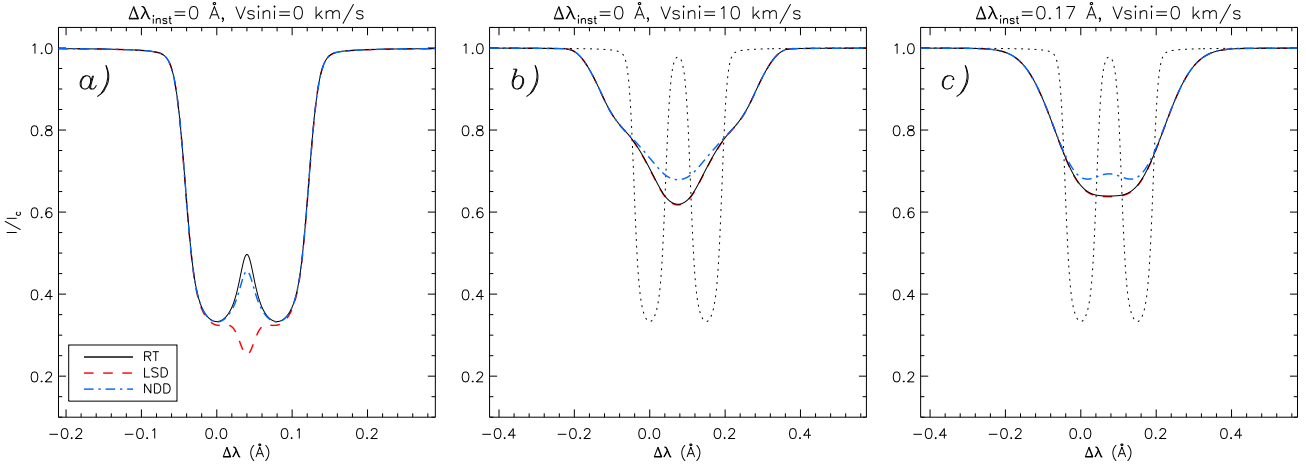}}}}
\caption{Comparison of the LSD ({\it dashed line}) and NDD ({\it dash-dotted line}) description of blended lines with the detailed radiative transfer (RT) calculation ({\it solid line}).
{\bf a)} Two identical unresolved lines with overlapping absorption coefficients. {\bf b)} Two resolved lines that overlap because of the rotational broadening corresponding to $v_{\rm e}\sin i$\,=\,10~\kms. {\bf c)} Same as {\bf b)}, but the overlap is caused by the Gaussian instrumental broadening with $FHWM=0.17$~\AA.}
\label{fig:ndd}
\end{figure*}

A simple modification of the LSD procedure allows circumventing the profile quality degradation associated with deconvolution. It also allows us to decouple the mean profile noise characteristics from the step size of the velocity grid on which the LSD profile is reconstructed. We modify Eq.~(\ref{eq:lsq}) to require
\beq
\chi^2 + \Lambda \mathcal{R} \to \mathrm{min},
\eeq
where $\mathcal{R}$ is the regularization function, penalizing large pixel-to-pixel variations of the LSD profile, and $\Lambda$ is the corresponding regularization parameter. The first-order Tikhonov regularization,
\beq
\mathcal{R} = \frac{\displaystyle \Lambda}{\displaystyle 2}\sum_i \left[ (Z_i - Z_{i-1})^2 + (Z_i - Z_{i+1})^2 \right],
\eeq
which minimizes the first line profile derivatives, is a convenient choice of regularization for this problem. It can be implemented in the LSD matrix algorithm by modifying Eq.~(\ref{eq:lsd}) as follows
\beq
\mathbf{Z} = (\mathbf{M}^T \cdot \mathbf{S}^2 \cdot \mathbf{M} + \Lambda \mathbf{R})^{-1} \cdot \mathbf{M}^T \cdot \mathbf{S}^2 \cdot \mathbf{Y}^{\rm o},
\label{eq:rlsd}
\eeq
where the tri-diagonal $m\times m$ matrix $\mathbf{R}$ is given by
\beq
\mathbf{R} = 
\left(\begin{array}{rrrr}
1 & -1 & 0 & \cdots \\
-1 & 2 & -1 & \cdots \\
\vdots & \multicolumn{2}{c}{\ddots} & \vdots \\
\cdots & -1 & 2 & -1 \\
\cdots & 0 & -1 & 1 \\
\end{array}\right).
\eeq

Figure~\ref{fig:reg} demonstrates that with an appropriate choice of the regularization parameter one can deconvolve the cross-correlation function without degrading its noise properties. In this particular example the regularized deconvolution yields roughly 50\% gain in S/N compared to the standard LSD technique.

The least-squares deconvolution constrained by the Tikhonov regularization of the mean profile is implemented in our code, but is not applied in the numerical experiments of Sect.~\ref{tests} because those are conducted for the noise-free spectra.

\subsubsection{Nonlinear deconvolution}

Several studies suggested modifications of the LSD method that would overcome its most important limitation: the assumption of linear addition of the overlapping lines, which leads to a poor description of the Stokes $I$ spectrum. The simplest way to reduce the discrepancy between the coarse LSD description and real intensity spectrum is to adjust not only the mean profile, but also the line strengths themselves, as done by \citet{reiners:2003} in their ``physical least-squares deconvolution''. This enhancement produces a superficially better fit to Stokes $I$. However, the mathematical formulation and numerical implementation of this method is somewhat ambiguous because one cannot determine the amplitude of the mean profile and line strengths in the LSD mask at the same time. One can also encounter uniqueness problems when fitting intensities for a group of very close lines. In addition, this technique cannot be easily extended to polarization analysis. In general, it yields the adjusted line strengths that are valid for Stokes $I$ but may be inappropriate for other Stokes parameters.

Recognizing that the main limitation of the classical LSD is its primitive treatment of blends, \citet{sennhauser:2009a} developed ``nonlinear deconvolution with deblending'' (NDD) technique in which the overlapping lines are added nonlinearly, using a formula derived from the \citet{minnaert:1935} approximation of the line shapes in the solar spectrum. Although \citet{sennhauser:2009a} argued that NDD is superior to LSD in the description of Stokes $I$, in their method they failed to distinguish two different types of blending present in the stellar spectrum. Very close spectral lines with overlapping absorption coefficients indeed add up nonlinearly. But the residual intensities of separated lines whose profiles overlap because of rotational, macroturbulent or instrumental broadening are simply added together, just like the standard LSD assumes. In Fig.~\ref{fig:ndd} we illustrate the difference between NDD and LSD description of blends in these two cases. Evidently the NDD method does not always yield a better approximation of the blended lines and in many cases appears to be inferior to LSD.

We agree that a nonlinear treatment of overlapping lines is a promising direction for improving LSD. However, one has to somehow distinguish the linear and nonlinear blending. One can envisage a successor to LSD that uses a method similar to NDD for the calculation of an unbroadened template spectrum and then uses the broadening function approach \citep{rucinski:1992} to account for various broadening effects that can be treated as convolution. We postpone the development and implementation of this two-stage multiline technique to future papers.

\subsection{Normalization of the LSD profiles}

A mathematical description of the stellar spectra by the LSD technique does not allow one to determine the mean profile amplitude independently of the scaling of weights in the line mask $\mathbf{M}$. Observations in each Stokes parameter constrain the product of the corresponding line weight and the LSD profile. Consequently, the mean profile amplitude depends on the arbitrary normalization of the line weights. This normalization determines parameters that should be assigned to the resulting mean profile. For example, the general form of the LSD line weights is
\beq
w_V = \frac{d \lambda \gbar}{d_0 \lambda_0 \gbar_0}
\label{wgtv}
\eeq
for Stokes V and
\beq
w_{QU} = \frac{d \lambda^2 \Gbar}{d_0 \lambda^2_0 \Gbar_0}
\eeq
for linear polarization. 

Normalization by the parameters $f_0=d_0, \lambda_0, \gbar_0$ and $\Gbar_0$ can be equivalently applied to the input LSD weights or to the resulting mean profiles. According to Sect.~4.1 of their paper, \citet{donati:1997} have arbitrarily chosen $\lambda_0=500$~nm, $d_0=1$ and $\gbar_0=1$. In other words, their reconstructed LSD profiles should correspond to a line with both the central intensity and the effective Land\'e factor set to unity. Most other studies employing the LSD technique neglected to explicitly describe the adopted normalization of the LSD weights, thus obscuring the interpretation of their published mean profiles.

On the other hand, many LSD-based investigations quote the mean line parameters, usually computed taking into account the statistical uncertainty of the observed spectrum:
\beq
\langle f \rangle = \frac{\sum_l f_l / \sigma^2_l}{\sum_l 1 / \sigma^2_l},
\label{eq:av1}
\eeq
where $f_l$ is the parameter that is averaged and $\sigma_l$ is an estimate of the observational uncertainty at the location of line $l$. Occasionally \citep[e.g.,][]{barnes:2000}, the averaging takes into account the line weight itself
\beq
\langle f \rangle_{w} = \frac{\sum_l f_l w_l/ \sigma^2_l}{\sum_l w_l / \sigma^2_l}.
\label{eq:av2}
\eeq

Compared to Eq.~(\ref{eq:av1}), this method yields systematically different average parameters because $f_l$ and $w_l$ are correlated. For example, for a sample of 8864 lines used for the numerical experiments in Sect.~\ref{tests} of our paper we find $\langle \gbar \rangle = 1.21$ and $\langle \gbar \rangle_{w} = 1.41$ for the weight $w_V$ computed with Eq.~(\ref{wgtv}). Most LSD studies do not comment on the averaging method they use.

Note that no matter which expression is adopted for computing the average line parameters, these quantities characterize only certain statistical properties of a set of lines included in the LSD mask, but should not be assigned to the LSD profiles themselves as implied by \citet{donati:1997} and many subsequent studies. The parameters corresponding to the LSD profiles are determined uniquely by the arbitrary normalization constants $f_0$, which are not necessarily the same as the average parameters computed with Eqs.~(\ref{eq:av1}) and (\ref{eq:av2}), unless a special effort is made to ensure $f_0 = \langle f \rangle$
or $f_0 = \langle f \rangle_w$, as done by some LSD studies (V. Petit, private communication). Of course, the latter approach becomes advantageous when one uses a spectrum synthesis code, trying to interpret the LSD profiles as Stokes parameters of a real spectral line (see Sect.~\ref{lsd_prof}).

The confusion between the average line parameters and the LSD weight normalization persists through many papers and has possibly led to erroneous interpretations of the mean profiles in some of these publications. We recommend that studies employing the LSD method should follow simple guidelines in order to better document application of this technique and make their results transparent and reproducible:
\begin{itemize}
\item provide values of the parameters $d_0$, $\lambda_0$, $\gbar_0$, and $\Gbar_0$ (if linear polarization is involved) adopted for normalizing the LSD weights or renormalizing the LSD profiles;
\item clarify which expression is used to compute the average line parameters;
\item most importantly, explain which meaning is assigned to the average quantities and state whether or not the presented LSD profiles are rescaled to ensure $f_0 = \langle f \rangle$.
\end{itemize}

\subsection{Interpretation of the LSD profiles}

The average intensity and polarization profiles provided by the LSD technique can be interpreted with various assumptions and levels of sophistication. In addition to merely using LSD as a qualitative magnetic field detection tool \citep{donati:1997}, we can distinguish the following main directions of utilizing them in the stellar physics studies:
\begin{enumerate}
\item One can infer characteristics of a star and its magnetic field through a direct comparison of the LSD profiles obtained self-consistently (i.e. using the same line lists and weights) from observations and from the detailed theoretical spectrum synthesis of the entire stellar flux spectrum in several Stokes parameters. Then the LSD profiles are used only for the inter-comparison of theory and observations but are not assigned any particular physical meaning. Obviously, this is the most robust method of treating the output of the least-squares deconvolution, but it is still far beyond the capabilities of the current computational resources and thus hasn't been applied in any real study. 
\item Taking into account that the LSD procedure is linear, we can use the theoretical spectrum synthesis of the entire stellar intensity spectrum to derive local LSD profiles for a range of temperatures, chemical abundances, magnetic fields strengths, and orientations. Interpolating within this theoretical local LSD line profile table we can obtain the model LSD flux profiles in one or several Stokes parameters and compare them with the corresponding LSD spectra inferred from observations. 

This analysis technique has not been implemented yet. However, some studies employed a simpler version of this method where the LSD profiles obtained from the observed flux spectra of sharp-lined benchmark stars of different \teff\ are assumed to represent local profiles corresponding to different regions on the surface of spotted rapidly rotating active stars \citep{barnes:2000}. This approach neglects the variation of line profiles with the limb angle and typically relies on a linear interpolation between only two template stars. An extension to the circular polarization LSD analysis is achieved within the framework of the weak field approximation, adopting derivatives of the LSD Stokes $I$ profiles of sharp-lined stars as the local circular polarization spectrum \citep{donati:1997b}.
\item Assuming that the LSD profiles behave as a real spectral line with average parameters, one can interpret the LSD spectra with standard theoretical tools. This approach is frequently used in recent Doppler imaging studies of the stellar surface structure and magnetic fields \citep[e.g.,][]{donati:2008b,morin:2008,folsom:2008}.
\item Finally, we can limit ourselves to the analysis of moments of LSD profiles that are assumed to behave similar to the corresponding moments of real spectral lines. In particular,
\begin{itemize}
\item The zeroth-order moment of Stokes $I$, equivalent width 
\beq
W_v = d_0 \int Z^I \mathrm{d}v,
\eeq
can be used to measure metallicity or abundances of individual elements. Note that in the expression for $W_v$ the integral of $Z^I$ is scaled by the central line intensity $d_0$ adopted for the normalization of the Stokes $I$ LSD weights.
\item The first-order moment of Stokes $V$ provides a measure of the mean longitudinal magnetic field
\beq
\langle B_{\rm z} \rangle = - 7.145\times 10^6 \frac{\int v Z^V \mathrm{d}v}{\lambda_0 \gbar_0 \int Z^I \mathrm{d}v}, 
\label{bz}
\eeq
where the magnetic field is measured in Gauss, the wavelength in \AA, and velocity in \kms. The wavelength $\lambda_0$ and effective Land\'e factor $\gbar_0$ appearing in this expression represent the arbitrary quantities adopted for the normalization of the Stokes $V$ LSD weights (see Eq.~(\ref{wgtv})), but not the mean line parameters, as often erroneously stated in LSD papers \citep{donati:1997,wade:2000}.
\item The normalized equivalent width of the LSD Stokes $Q$ and $U$ LSD profiles
\beq
P_Q = \frac{\langle w_{QU} \rangle \int Z^Q \mathrm{d}v}{\int Z^I \mathrm{d}v} \mathrm{~~and~~}
P_U = \frac{\langle w_{QU} \rangle \int Z^U \mathrm{d}v}{\int Z^I \mathrm{d}v},
\eeq
where $\langle w_{QU} \rangle$ is the average Stokes $QU$ LSD weight, characterizes the net linear polarization \citep{wade:2000} and has a similar information content as the broad-band linear polarization measured with the help of photopolarimetric devices \citep{leroy:1995a}.
\end{itemize}
\end{enumerate}

It is beyond the scope of this paper to explore all these methods of interpreting the LSD profiles. The numerical experiments presented below are mainly focused on testing the third approach and several aspects of the fourth approach.

\section{Numerical tests of the LSD technique}
\label{tests}

\subsection{Polarized spectrum synthesis}
\label{synt}

\begin{figure*}[!t]
\firps{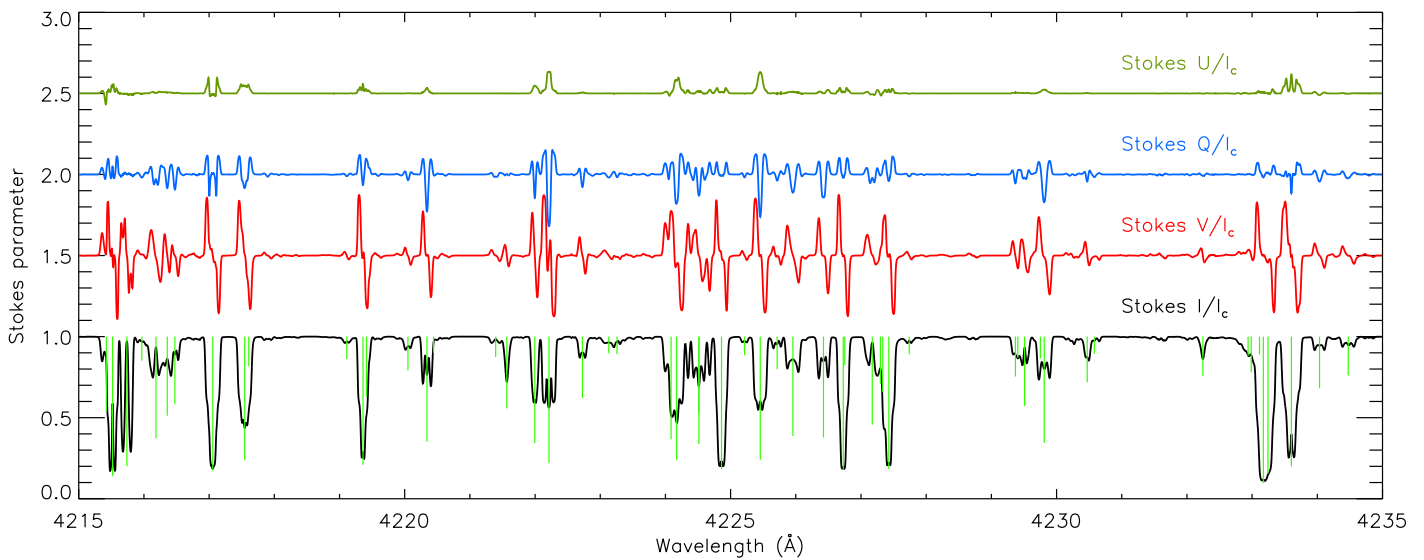}
\caption{Small segment of the normalized local four Stokes parameter spectra calculated with \syn\ for the disk center ($\mu=1$), field strength $B=5000$~G and magnetic vector orientation $\gamma=45\degr$, $\chi=0\degr$. Linear and circular polarization profiles are shifted vertically for display purpose. Thin vertical bars plotted over the Stokes $I$ spectrum indicate absorption lines deeper than 10\% in the absence of magnetic field.}
\label{fig:spec}
\end{figure*}

We use the magnetic spectrum synthesis code \syn\ to calculate four Stokes parameter stellar spectra covering a large part of the optical wavelength range. \syn\ is a further development of the {\sc synthmag} code, described in detail by \citet{piskunov:1999} and \citet{kochukhov:2007d}. We refer the readers to these papers for an in-depth discussion of the numerical techniques and input physics implemented in the code. In summary, we solve numerically the polarized radiative transfer equation for a given stellar model atmosphere and homogeneous or vertically stratified chemical abundances assuming local thermodynamical equilibrium. The code includes an updated version of the molecular equilibrium solver developed by \citet{valenti:1998}, allowing us to treat Zeeman and partial Paschen-Back splitting in both atomic and molecular absorption lines. 

Previous studies extensively used {\sc synthmag} calculations of individual spectral lines or blends in narrow wavelength intervals to investigate the magnetic field, chemical composition, and stratification in A and B chemically peculiar stars \citep[e.g.,][]{kochukhov:2003a,kochukhov:2006c,ryabchikova:2006} and to model magnetic fields of T~Tauri stars \citep{johns-krull:2007} and active M dwarfs \citep{johns-krull:1996,kochukhov:2009b}. In the present study we have extended the capabilities of the code, adapting it for massive polarized spectrum synthesis of thousands of spectral lines.

The output of \syn\ is a set of local Stokes $IQUV$ spectra for a given magnetic field vector\footnote{The magnetic field vector is characterized by the field modulus $B$, angle $\gamma$ between the field vector and the line of sight and angle $\chi$, giving orientation of the field in the plane perpendicular to the line of sight. Figure~2 in \citet{piskunov:2002a} illustrates the definition of these angles.} and the angle of view from the normal to the surface ($\mu\equiv\cos\theta$). These spectra can be combined with appropriate weights and Doppler shifts to simulate stellar flux spectra for a simple homogeneous magnetic field distribution \citep{kochukhov:2007d} or for an arbitrary prescribed map of surface structures \citep{piskunov:2002a}. However, because each of these flux spectra represents a linear superposition of the local intensity spectra computed for different magnetic field parameters and the LSD procedure behaves as a linear operator, it is sufficient to limit the investigation to the local, disk-center Stokes parameters that are calculated for a sufficiently broad range of the field strengths and inclinations. This analysis should provide a basic but still rather general assessment of the performance of the LSD technique without digressing to the analysis of particular surface magnetic field configurations.

For the magnetic and chemical abundance LSD experiments presented in this section we considered an {\sc atlas9} \citep{kurucz:1993c} model atmosphere of the star with \teff\,=\,9000~K and \logg\,=\,4.0. We adopted solar abundances for all elements except Cr and Fe. The concentration of these two species was enhanced by 2 and 1~dex, respectively, which is typical of the magnetic chemically peculiar stars in this temperature range \citep{ryabchikova:2005b}. We used the VALD database \citep{kupka:1999} to compile a list of 8864 metal lines with the residual intensity exceeding 1\% of the continuum in the 4000--6000~\AA\ wavelength region. The broad hydrogen Balmer lines and weak \ion{He}{i} features were excluded from the line list. The Zeeman splitting patterns required for the polarized spectrum synthesis and calculation of the LSD weights were obtained using Land\'e factors provided by VALD when available and estimated using the LS-coupling scheme otherwise.

Theoretical spectra in four Stokes parameters were calculated on the adaptive wavelength grid \citep[see][]{kochukhov:2007d} and then rebinned to an equidistant velocity spacing of 0.62~\kms\ and convolved with Gaussian profiles of $FWHM$\,=\,2.5~\kms. This sampling and resolution corresponds to the spectra provided by the HARPSpol instrument \citep{snik:2008} at the ESO 3.6-m telescope, which is the highest resolution stellar spectropolarimeter currently in operation. A few additional tests were done with the reduced resolving power of 65\,000 ($FWHM$\,=\,4.6~\kms), corresponding to the polarization spectra from the ESPaDOnS and NARVAL instruments. 

Figure~\ref{fig:spec} shows a small segment of the local four Stokes parameter profiles calculated with the help of \syn.

\subsection{Self-similarity of spectral lines}

First we examine the hypothesis of self-similarity of the Stokes profiles of different spectral lines, which is a fundamental assumption made by the LSD technique. We have selected all \ion{Fe}{i} and \ion{Fe}{ii} lines from the original line list, removing a few iron lines with a very weak magnetic field sensitivity. The resulting 1652 lines of diverse strengths and Zeeman patterns were redistributed in wavelength to avoid blending. This line list was employed for the calculation of a grid of local Stokes parameter spectra for the field strength from 0.1 to 5~kG and the field inclination from 0 to 90\degr. The spectra were then convolved with the Gaussian instrumental profiles.

A quantitative measure of the line self-similarity was obtained by computing the standard deviation of the line profiles interpolated onto a common velocity grid and divided by the appropriate LSD weights. We tested the standard LSD weights discussed in Sect.~\ref{lsd} as well as alternative weights obtained by substituting the central line depth $d$ with the equivalent width expressed in velocity units, $W_v$.

\begin{figure*}[!t]
\centering
\resizebox{16cm}{!}{\rotatebox{90}{\includegraphics{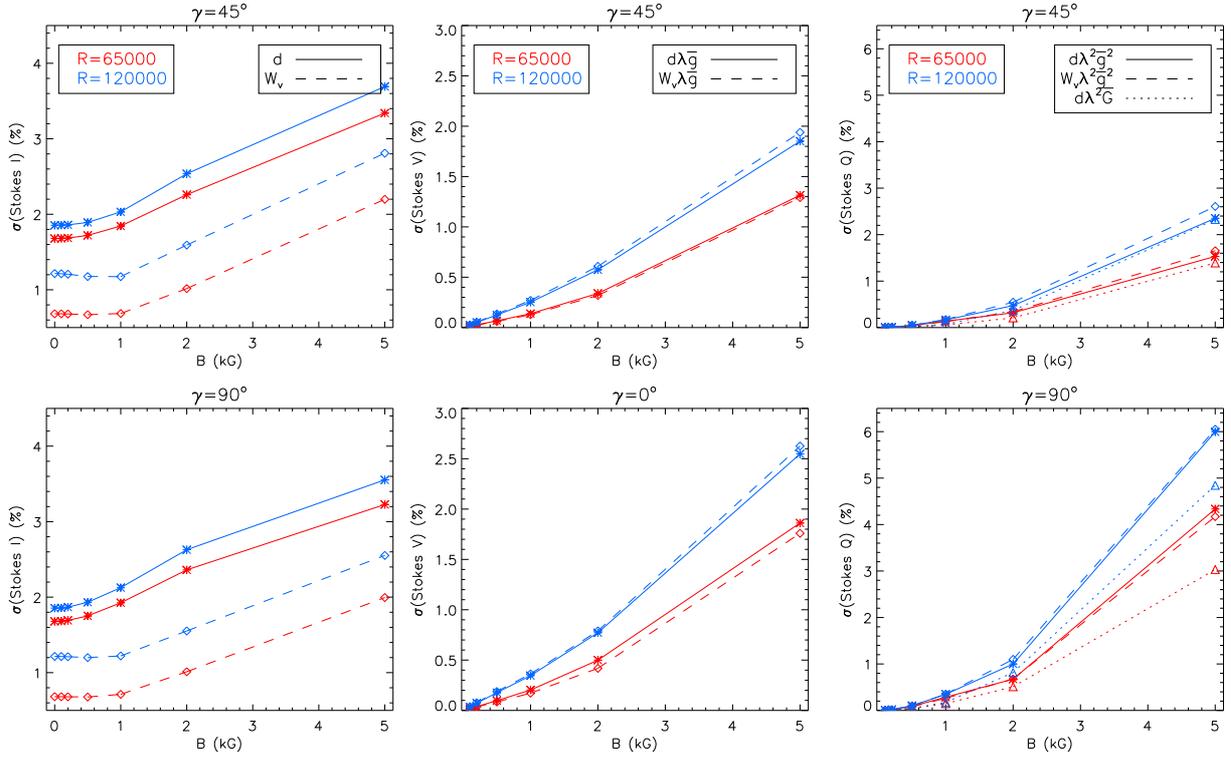}}}
\caption{Standard deviation of the scaled Stokes $I$, $V$ and $Q$ profiles as a function of the field strength for two different values of the field inclination $\gamma$ (45\degr, 90\degr\ for $I$ and $Q$ and 45\degr, 90\degr\ for $V$). Results for the two resolving powers are shown with red and blue curves. The solid, dashed, and dotted lines correspond to the standard deviations obtained using different line scaling factors.}
\label{fig:scale}
\end{figure*}

The standard deviation of the rescaled Stokes profiles is plotted as a function of the field strength in Fig.~\ref{fig:scale}. This figure shows results for two different resolutions. Clearly for Stokes $I$ the line-self similarity assumption is a poor one and might be appropriate only for a comparatively low $S/N$ of the observational data. Scaling by the equivalent width instead of the central line depth evidently gives better results. This reflects that strong lines are also systematically broader and therefore their central intensity is less affected by additional broadening. Based on this test we recommend to use the equivalent width rather than the line depth as an LSD weight in the applications concerned with Stokes $I$ alone. However, the relative strength of the polarization signatures is described reasonably well by both $d$ and $W_v$.

\begin{figure*}[!t]
\centering
\resizebox{17cm}{!}{\rotatebox{90}{\includegraphics{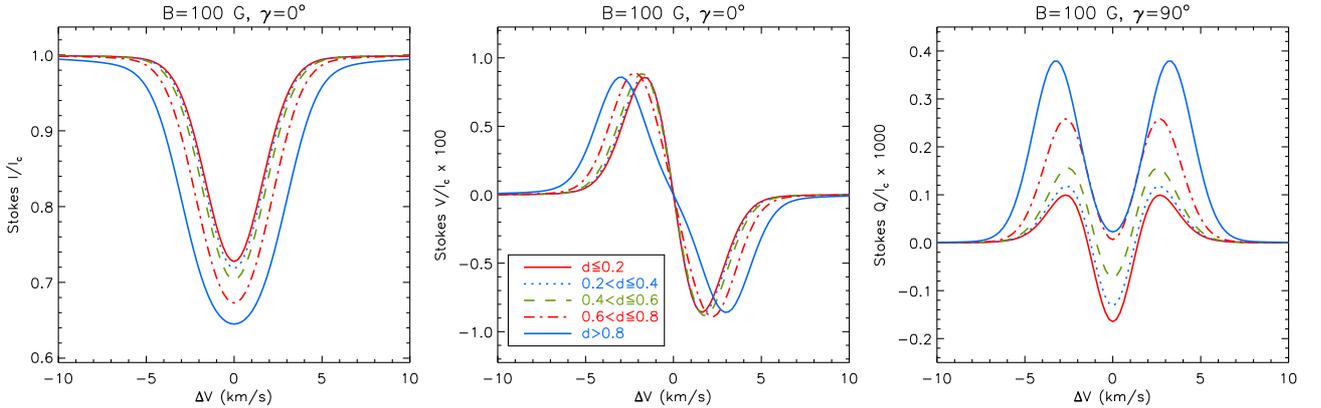}}}
\caption{Mean scaled Stokes $I$, $V$, and $Q$ signatures for the Fe lines of different central intensity. These profiles are obtained from the spectra computed with $B=100$~G and an inclination angle $\gamma=0\degr$ (Stokes $I$ and $V$) and $\gamma=90\degr$ (Stokes $Q$).}
\label{fig:b100}
\end{figure*}

As is evident from Fig.~\ref{fig:scale}, the assumption of the similarity of the line profile shapes is a reasonable one for Stokes $V$ if the field strength does not exceed $\sim$\,2~kG and the observational noise is $\ge 5\times10^{-3}$ (which corresponds to $S/N=200$). Obviously, as the resolution and $S/N$ increases, we see more and more difference between individual spectral lines. The diversity of the linear polarization profiles is far greater than that of Stokes $V$. Even for the field strength of $\sim$\,1~kG profiles noticeably differ from each other, and this difference rapidly increases with the field strength. Figure~\ref{fig:scale} shows that a definite reduction of the scatter between the scaled Stokes $Q$ profiles is achieved by using $\Gbar$ instead of $\gbar^2$ in the corresponding line weight.

The reason for the greater diversity of the Stokes $Q$ line profiles lies in their sensitivity to the line strength. As emphasized in Sect.~\ref{polar}, the linear polarization profiles are expected to be similar only when both the weak-field \textit{and} the weak-line assumptions are satisfied, while for circular polarization only the former approximation is directly necessary. It is instructive to illustrate this by examining the average Stokes profiles for a very low field strength, when the weak-field approximation is valid. Figure~\ref{fig:b100} shows that in this situation there is some difference between the Stokes $I$ line profile shapes of the weak and strong lines (as already discussed above) and a considerable difference between the shapes of $Q$ profiles. The shape of the average linear polarization signature changes smoothly with the line strength, suggesting that the latter is the primary reason for the systematic discrepancy of the individual Stokes $Q$ line profiles. In contrast, the Stokes $V$ profiles are all very similar, almost independently of the line strength. Consequently, the basic assumptions of the LSD method are fairly robust for Stokes $V$ but are more questionable for linear polarization.

\subsection{The cutoff criterium and LSD profiles of individual elements}

In the second set of numerical experiments, conducted for the local non-magnetic spectrum, we investigated the effect of choosing a different cutoff criterium for the selection of lines that contribute to the LSD mask. Previous studies based on the LSD technique arbitrarily applied a wide range of criteria: from a central depth of 0.1 to 0.4. We found that for our noise-free spectra increasing the cutoff leads to the downward displacement of the resulting mean profile relative to the continuum, presumably because of the combined effect of numerous weak lines that are excluded from the mask when the cutoff is too high. This continuum depression of the LSD Stokes $I$ profiles introduces an additional uncertainty in their interpretation. In subsequent tests we include in the mask all lines deeper than 10\% in the absence of magnetic field, giving us a total of 3270 usable lines.

\begin{figure}[!t]
\fifps{7.5cm}{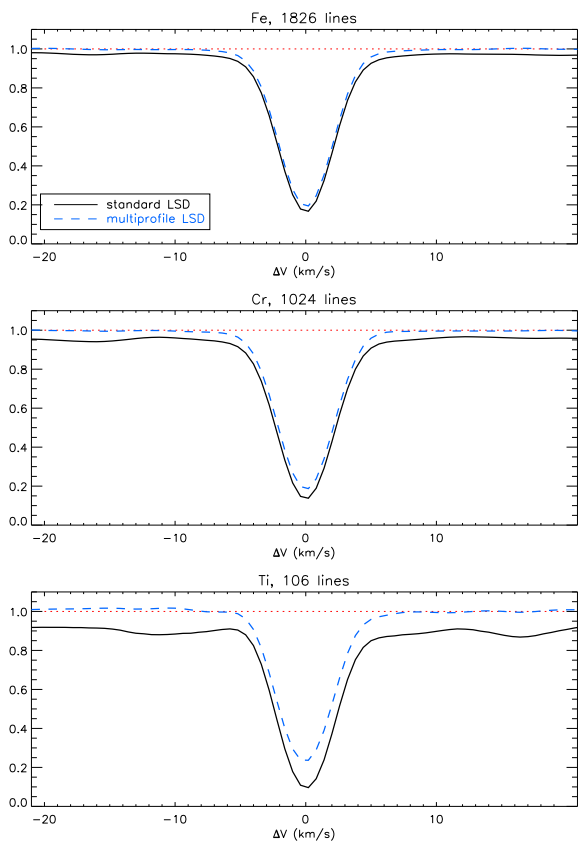}
\caption{Stokes $I$ LSD profiles of individual elements reconstructed with the standard LSD technique (solid lines) and multiprofile LSD (dashed lines).}
\label{fig:mlsd}
\end{figure}

We have also assessed the performance of our multiprofile extension of the LSD technique compared to the standard decomposition procedure. Using the non-magnetic spectrum we calculated the LSD profiles for Fe, Cr, and Ti in one case using only the lines of each individual element and in another case accounting also for all other lines with the help of the second, ``background'' LSD profile. Figure~\ref{fig:mlsd} presents a comparison of the LSD profiles for the three elements obtained with these two varieties of the multiline technique. One can see that our multiprofile version of LSD is able to recover more reliable mean line profiles, free of the distortions caused by blends and without the spurious continuum offset. This improvement is most prominent for Ti, which has only 106 useful lines in our synthetic spectrum. We therefore suggest the use of the multiprofile LSD in studies concerned with the analysis of small groups of lines present in stellar spectra.

\subsection{Longitudinal magnetic field}
\label{lsd_bz}

\begin{figure}[!t]
\figps{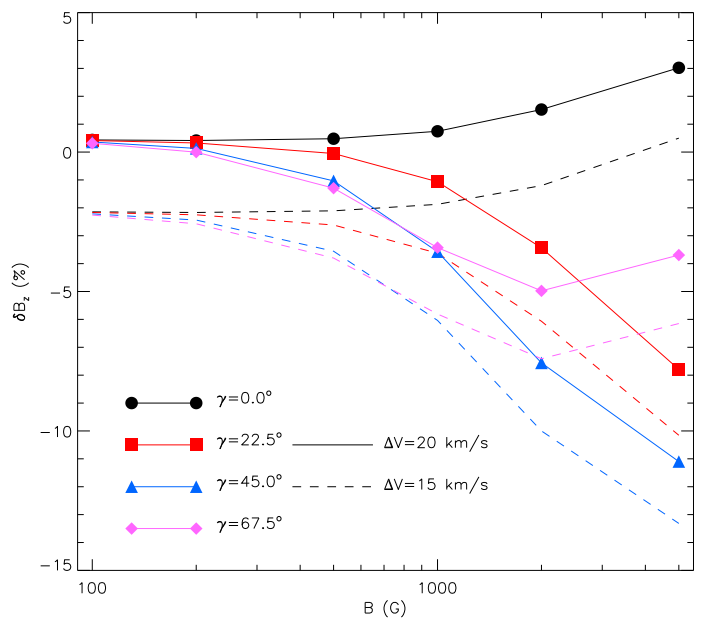}
\caption{Relative error of the longitudinal magnetic field $B_{\rm z}$ determined from the LSD Stokes $V$ profiles. The lines of different colors show results for different inclinations of the magnetic field vector, while the solid and dashed lines illustrate the effect of choosing different integration limits.}
\label{fig:bz}
\end{figure}

In the next series of tests we studied how accurately the line of sight magnetic field component $B_{\rm z}$ (longitudinal field) can be recovered from the LSD Stokes $V$ profiles. We computed a set of 30 local polarized spectra, including spectral lines of all chemical elements, for a grid of magnetic field strengths ($B=100$, 200, 500, 1000, 2000, 5000 G) and the angles between the field vector and the line of sight ($\gamma=0\degr$, 22.5\degr, 45\degr, 67.5\degr, 90\degr). For each of these theoretical spectra we obtained the mean Stokes $V$ profile and estimated the longitudinal field from its first moment, using Eq.~(\ref{bz}).

Figure~\ref{fig:bz} illustrates the relative error of $B_{\rm z}$ as a function of the field strength and inclination. We find that the analysis of the LSD profiles recovers the true value of the longitudinal field with an accuracy of a few percent for the fields below 1~kG. At the same time, as is evident from Fig.~\ref{fig:bz}, there is an uncertainty in the choice of the measurement window, resulting in another $\approx$\,2\% error of $B_{\rm z}$. 

The longitudinal field estimate becomes less accurate when the field strength exceeds 1~kG. We find that for small $\gamma$ angles the LSD profile analysis overestimates the longitudinal field by up to 3\%, while $B_{\rm z}$ can be underestimated by as much as 10\% for the intermediate magnetic field inclinations.

We conclude that the derivation of the longitudinal magnetic field from LSD profiles appears to be quite robust but can be affected by the systematic errors of the order of a few percent. For strongly magnetic stars the longitudinal field tends to be underestimated by the LSD technique.

\subsection{LSD profiles in Stokes parameters}
\label{lsd_prof}

\begin{figure*}[!t]
\firps{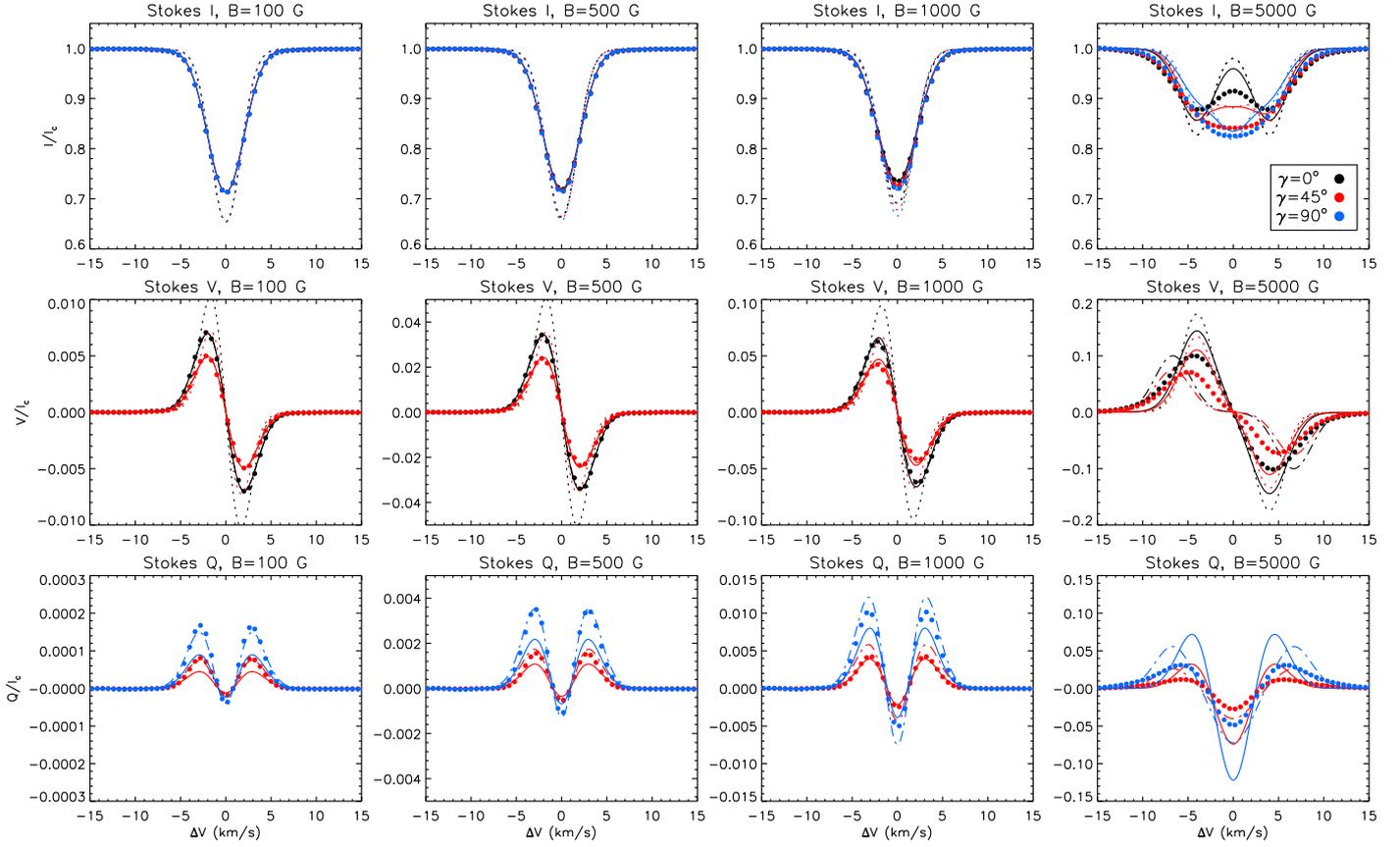}
\caption{Comparison of the LSD Stokes $IQV$ profiles obtained from synthetic stellar spectra (symbols) with the theoretical line profiles computed for the average line parameters (lines). Different colors illustrate the results for three values of the magnetic field inclination with respect to the line of sight (0\degr, 45\degr\ and 90\degr). The four columns correspond to different magnetic field strength values (0.1, 0.2, 1 and 5 kG). The solid lines show theoretical profiles for the resolution $R=80\,000$.
The dotted lines in the Stokes $I$ and $V$ panels illustrate calculations for $R=120\,000$, which was the instrumental broadening applied to the original simulated stellar spectra. The dash-dotted lines in the Stokes $V$ and $Q$ panels represent calculations for the equivalent Zeeman splitting pattern with widely separated $\sigma$ components.}
\label{fig:line}
\end{figure*}

Another series of numerical experiments was carried out to answer the question whether or not the LSD Stokes parameter spectra behave as a real spectral line with respect to changes of the magnetic field strength and orientation. We derived LSD profiles from a grid of synthetic spectra covering the same parameter space as studied in Sect.~\ref{lsd_bz}. The mask adopted in the least-squares deconvolution procedure was normalized using $\langle \lambda \rangle\approx\lambda_0=4700$~\AA, $d_0=\langle d \rangle=0.36$ and $\gbar_0=\langle\gbar\rangle=1.22$. The resulting mean Stokes $I$, $V$ and $Q$ LSD profiles were compared with the theoretical spectra calculated for a fictitious \ion{Fe}{i} line with the central wavelength $\lambda=4700$~\AA, an excitation potential of the lower level (3.47 eV) matching the average for this ion and a normal Zeeman triplet splitting pattern characterized by $\gbar=1.22$. The oscillator strength of the \ion{Fe}{i} line was adjusted to reproduce the equivalent width of the Stokes $I$ LSD profile in the absence of magnetic field.

The comparison of the LSD profiles and synthetic single-line calculations is presented in Fig.~\ref{fig:line} for a subset of the considered magnetic field parameters. We found that even for the non-magnetic case the synthetic calculations for the resolving power $R=120\,000$ (dotted lines in Fig.~\ref{fig:line}) do not match the LSD Stokes $I$ line profile shape reconstructed from the full synthetic spectra having this resolution. The spectral resolution had to be decreased to approximately $R=80\,000$ (solid lines in Fig.~\ref{fig:line}) to reproduce the weak-field LSD profiles. This is probably an effect of the extended wings of strong lines, which contribute disproportionally to the mean profile. Thus, an estimate of the macroscopic velocity fields (stellar rotation, macroturbulent velocity) using LSD profiles might be affected by a small bias, especially for cooler stars in which the strong lines with developed damping wings are considerably more important than for the $T_{\rm eff}=9000$~K stellar model considered here.

Considering the behavior of the LSD Stokes $I$ spectrum with the field strength, we found a fair agreement with the theoretical single-line calculations up to $B\approx1$~kG. Beyond this field strength value the lines start to exhibit partially resolved Zeeman patterns and the discrepancy between the LSD profile and its theoretical single-line approximation becomes progressively worse.

The circular polarization profiles show qualitatively the same picture as Stokes $I$. Below the field strength of $\sim1$~kG there is an excellent agreement between theoretical Stokes $V$ line shape and the Stokes $V$ LSD profile, provided the extra Gaussian broadening is applied to the single-line calculations. The agreement becomes poor for $B\ge2$~kG when individual Zeeman splitting patterns of spectral lines start to influence their Stokes $V$ profiles. For strong fields, illustrated by the $B=5$~kG profiles in Fig.~\ref{fig:line}, the LSD Stokes $V$ spectra are typically broader and have a lower amplitude compared to the single-line calculations. The discrepancy between LSD profiles and single-line calculations shows no clear dependence on the field orientation.

In contrast to the encouraging weak-field behavior of the Stokes $I$ and $V$, the Stokes $Q$ single-line calculations fail in reproducing the corresponding LSD profiles for all field strengths. The shapes of the liner polarization signatures agree well for lower field strengths. But the amplitude of the LSD Stokes $Q$ spectrum is systematically underestimated by the single-line synthesis for the fields weaker than $\sim1$~kG. Then, as the field increases, the single-line Stokes $Q$ amplitude grows faster than that of the LSD profile. The two amplitudes roughly agree for $B\approx1.5$~kG and disagree in the other sense for stronger fields. This prominent field strength dependence implies that the disagreement that we revealed cannot be removed by a simple, field-strength-independent re-scaling of the LSD Stokes $Q$ profiles.

The outcome of the calculations presented in this section suggests that in general the LSD profiles in four Stokes parameters do not behave as a single spectral line with a triplet Zeeman splitting pattern. It appears that the effective Land\'e factor and the line strength that successfully reproduce the LSD Stokes $I$ and $V$ spectra do not match the amplitude of LSD Stokes $Q$ even for weak fields. There is also a  significant systematic disagreement between the shapes of the LSD Stokes $I$ and $V$ profiles and single-line theoretical calculations for the field strengths above $\approx2$~kG.

In reality the majority of spectral lines exhibit anomalous Zeeman splitting, which can be considerably more complicated than a pure triplet. In principle, the positions and strengths of individual components within the groups of red- and blue-shifted $\sigma$ components can be manipulated to change the relative amplitude of the Stokes $V$ and $Q$ profiles, potentially providing a somewhat better agreement between the LSD polarization profiles and single-line calculations. For example, the dash-dotted lines in Fig.~\ref{fig:line} illustrate the Stokes $Q$ profiles for a splitting pattern with widely separated pairs of $\sigma_+$ and $\sigma_-$ components. The effective Land\'e factor is kept at the value of 1.22 and neither the amplitude nor the shape of Stokes $V$ change compared to the magnetic spectrum synthesis for the triplet splitting unless the field is very strong. On the other hand, the amplitude of the Stokes $Q$ increases, providing a better match to the LSD Stokes $Q$ profiles for a limited field strength range. However, parameters of this splitting pattern cannot be inferred from any properties of the LSD line mask and have to be calibrated empirically for a prescribed field strength range.

\subsection{Abundance determination from LSD profiles}

In this part of our investigation we assess to what extent LSD profiles behave as a real spectral line with respect to changes of abundances of individual elements. Specifically, we want to find out if there exists a set of atomic parameters reproducing changes of the LSD profile with abundance if it is modeled as a single spectral line. For the Ap-star like abundances adopted for our numerical tests the spectrum is dominated by the lines of Fe and Cr. Hence we considered the effects of the abundance variation for these two elements. 

Our LSD code allows to reconstruct an arbitrary number of independent LSD profiles using several masks. We took advantage of this feature in the abundance tests, reconstructing two LSD profiles: one for the element of interest and another for all other lines in the mask.

Assuming the iron abundance [Fe]=+1 dex (hereafter abundances are given in the logarithmic scale with respect to the Sun), five non-magnetic theoretical spectra were calculated for various abundances of Cr in the range from $+1$~dex to $+3$~dex with a step of 0.5~dex. Then the chromium abundance was kept at [Cr]=+2~dex, while the abundance of Fe was changed from 0~dex to $+2$~dex. This gave us five spectra with different Cr and five calculations with different Fe abundance, with a common spectrum for [Fe]=+1~dex and [Cr]=+2~dex. The latter spectrum was used for the selection of lines included in the LSD mask using our standard 10\% cutoff criterium for the central line depth.

\begin{figure}[!t]
\centering
\firps{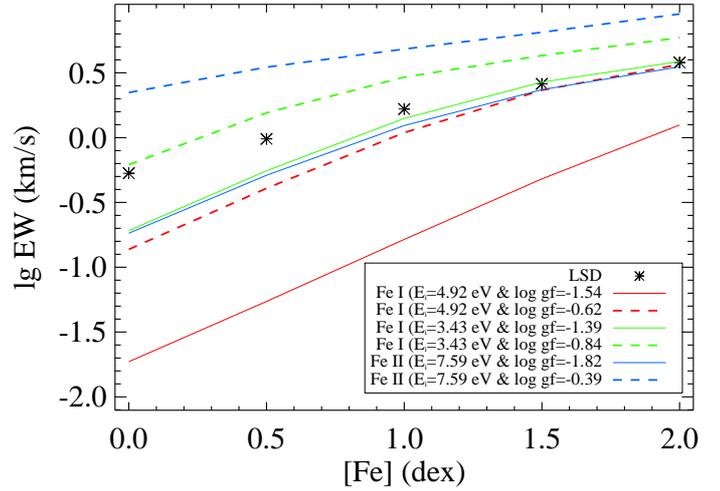}
\caption{Dependence of the equivalent width on abundance for the LSD profiles (\textit{symbols}) and theoretical profiles of Fe spectral lines (\textit{lines}). Different lines illustrate the EW-abundance dependence obtained for different combinations of line parameters. Red: averaging over all Fe lines; green: averaging over \ion{Fe}{i} lines; blue: averaging over \ion{Fe}{ii} lines. The dashed and solid lines show results for the oscillator strength determined using harmonic and arithmetic averaging, respectively.}
\label{fig:fe_ew}
\end{figure}

\subsubsection{Equivalent width}

The dependence of the equivalent width (EW) of the Fe LSD profile on the abundance of this element is illustrated in Fig.~\ref{fig:fe_ew}. We tried reproducing this dependence with single-line calculations using different methods of determining the parameters of the average line. Both the \ion{Fe}{i} and \ion{Fe}{ii} lines contribute at a similar level to our synthetic spectra and to the resulting LSD profiles. Hence one can adopt mean line parameters by averaging over the lines of particular ion or averaging over both ions. Furthermore, one can determine the mean oscillator strength using the harmonic or arithmetic average, the latter giving systematically higher value of the oscillator strength. This yields six possible combinations of ion, excitation potential, and oscillator strength. It is not clear which one should be chosen for modeling LSD profiles, hence we tested all of them. The resulting equivalent width changes with abundance are illustrated with the lines in Fig.~\ref{fig:fe_ew}. Evidently, two or three combinations of the mean line parameters give somewhat better results, but all predict much steeper slopes of the EW dependence on abundance than shown by the LSD profile. 

Calculations for Cr show a very similar discrepancy and reveal the same pair of the line parameters producing a slightly better fit to the LSD EWs compared to other options.

\begin{figure}[!t]
\centering
\firps{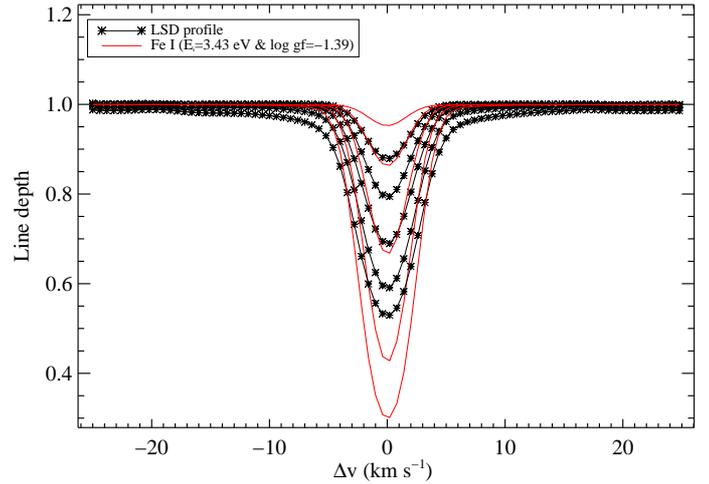}
\caption{Comparison of the Fe LSD profiles (\textit{symbols}) and the profiles of theoretical spectral line (\textit{solid curve}) for one of the sets of atomic parameters used in Fig.~\ref{fig:fe_ew}.}
\label{fig:fe_line}
\end{figure}

\subsubsection{Line profiles}

We complement the EW comparison with a more detailed investigation of the Fe LSD profiles as a function of abundance. This could potentially shed some light on the discrepancy between the LSD and expected single-line dependence on element abundance.

As shown in Fig.~\ref{fig:fe_line}, the LSD profiles appear to be stronger than the synthetic spectrum for lower abundances and weaker, but wider, for larger abundances. These broad wings do not contribute significantly to EW, and their exclusion would only degrade even more the agreement between the LSD profile and single-line abundance behavior.

We have verified that blending of the Fe or Cr lines by the lines of other elements is not the reason for the observed abundance dependence. In particular, if we calculate the pure Fe or Cr line spectrum and de-blend all lines by shifting them in wavelength, we still get the same slope of the EW vs. abundance trend as illustrated in Fig.~\ref{fig:fe_ew}.

We conclude that weak and strong lines are combined by the LSD technique in a way that the resulting mean Stokes $I$ profile does not show the abundance dependence of a real spectral line. This is probably because of a highly non-linear change of both the central depth and the width of stronger absorption lines with abundance variation. This suggests that deducing the absolute elemental abundances from LSD profiles is a highly non-trivial task, which can be accomplished only using LSD profile calculations from synthetic spectra similar to those presented here. Nevertheless, LSD profiles can still be used for a comparison of the element concentrations between different stars with similar fundamental parameters or different regions on the surface of the same star.

\section{Conclusions}
\label{conc}

We have carried out a detailed study of the least-squares deconvolution of both intensity and polarization stellar spectra. This multiline signal-enhancing technique is widely used in a broad range of modern spectroscopic and spectropolarimetric studies, but lacks a thorough  justification and was never tested using synthetic spectra. Our study fills this gap, clarifying the basic assumptions of LSD, investigating various ways to improve its numerical implementation and, finally, testing the sanity of the physical information derived from LSD spectra.

The main conclusion of our study can be summarized as follows.
\begin{enumerate}
\item The basic LSD assumptions of line profile self-similarity and linear addition of blends are satisfied only for very weak (less than 20\% of the continuum) absorption lines in Stokes $I$. Similar results are obtained for linear polarization signatures, which exhibit a prominent line-strength dependence in weak fields and become quite diverse in shape owing to different Zeeman splitting patterns for $\ga$1~kG magnetic fields. On the other hand, the LSD assumptions are appropriate for the majority of lines in Stokes $V$ for weak and moderately strong magnetic fields ($B\le2$~kG).
\item We developed a new, improved LSD code, which is generalized to handle multiple mean profile reconstruction from the same spectrum and employs regularization to suppress excessive numerical noise that appears in the LSD profiles as a result of the autocorrelation matrix inversion. At the same time, we demonstrated that the recently suggested non-linear deconvolution method of \citet{sennhauser:2009a} is applicable only to the spectra of very slowly rotating stars observed at the highest available spectroscopic resolutions, but is inferior to the standard LSD method when the line blending is primarily due to rotational and instrumental broadening. 
\item We emphasized the importance of choosing correct line weights in the calculation of LSD profiles. It is the normalization of these weights, but not the mean line parameters, that determine the characteristics (central wavelength, Land\'e factor, etc.) of the resulting LSD profile. We suggested an improved weighting scheme for the computation of the linear polarization LSD profiles, which accounts for the anomalous Zeeman splitting of spectral lines.
\item We performed comprehensive numerical experiments in which we analyzed LSD profiles computed from the local theoretical Stokes parameter spectra of a mid-A magnetic chemically peculiar star. The measurements of the mean longitudinal magnetic field from the synthetic LSD profiles show that $B_{\rm z} $ accurate to within a few \% can be retrieved for the magnetic field strength of up to $\sim$1~kG, but systematic errors of up to 10\% can be expected for stronger fields. 
\item Generally, the LSD intensity and polarization profiles do not behave as a spectral line with parameters determined by averaging over individual lines contributing to the mean spectrum. In particular, the LSD Stokes $I$ profiles appears broader and its equivalent width changes to a lesser extent with variation of element abundances than for a real spectral line. The LSD Stokes $V$ and $Q$ cannot be described simultaneously by the same line parameters even in a weak magnetic field. The only parameter region where we found the single-line approximation of the LSD profiles to be reasonably robust is the circular polarization spectra in the 0--2~kG field strength interval. Whenever the field intensity exceeds this limit or an element abundance is changed or Stokes $Q$ is considered simultaneously with the circular polarization, the LSD profile cannot be predicted by a single-line spectrum synthesis. These results suggest that the single-line approximation of LSD profiles should be abandoned in favor of a direct comparison of the mean profiles inferred from observations and from the same lines in theoretical polarized synthetic spectrum.
\end{enumerate}

\begin{acknowledgements}
We thank V\'eronique Petit and Gregg Wade for valuable discussions of the LSD method.
O.K. is a Royal Swedish Academy of Sciences Research Fellow supported by grants from the Knut and Alice Wallenberg Foundation and the Swedish Research Council.
\end{acknowledgements}


\end{document}